\documentclass[journal, final]{IEEEtran}
\usepackage{subfigure}
\usepackage[dvips]{graphicx}
\usepackage{amsmath}
\usepackage{cite}
\usepackage{color}
\usepackage{graphicx}

\usepackage{ifpdf}
\ifpdf
\begin{document}

\title{Narrowband Bandpass Frequency Selective Surface with Miniaturized Elements}

\author{Amir~Ebrahimi,~\IEEEmembership{Member,~IEEE,} 
        Thomas~Baum~\IEEEmembership{Member,~IEEE,}\\
         James~Scott~\IEEEmembership{Member,~IEEE,} and Kamran~Ghorbani~\IEEEmembership{Member,~IEEE,}

        \it{
        \IEEEauthorrefmark{0}School of Engineering, Royal Melbourne Institute of Technology (RMIT University), VIC~3001, Australia}\\
                \rm amir.ebrahimi@rmit.edu.au\\

} %

\maketitle

\vspace{-5mm}

\begin{abstract}
This article presents a bandpass frequency selective surface (FSS) with a narrowband frequency response. The designed FSS is made of miniaturized elements unit cell. The operation principle of the FSS is explained by using an equivalent circuit model, where the passband bandwidth can be controlled by the values of the circuit elements corresponding to the geometrical dimensions of the unit cell. The compatibility of the presented structure in designing higher order narrowband filtering responses is verified by designing a second-order bandpass FSS with $8.5\%$ fractional bandwidth with a center frequency of $2.7$~GHz. 
\end{abstract}
\IEEEpeerreviewmaketitle
\vspace{-2mm}
\section{Introduction}
Frequency selective surfaces (FSSs) are artificial periodic structures that can manipulate the propagation of the electromagnetic waves by controlling their amplitude, phase, or polarization \cite{Abadi2015}. Due to their versatile functionalities, the FSSs have found a wide range of applications from low microwave to terahertz and optical frequencies \cite{Ebrahimi2015}. For example, they can be designed as absorbers \cite{Ebrahimi2016}, reflect arrays \cite{Niu2014}, transmit arrays \cite{Li2013b}, polarization convertors \cite{Joyal2012,Abadi2016}, and etc. A major application of the FSSs is in spatial filtering of the electromagnetic radiations for reducing the radar cross-section (RCS) of the objects in stealth platforms, reducing the interferences in communications, and shielding the electronic devices against unwanted signals \cite{Abadi2015}. Variety of the FSS-based filtering structures have been designed so far with different characteritics such as higher-order bandpass filtering \cite{Al-Joumayly2010}, elliptical or quasi elliptical filtering \cite{Li2013a,Rashid2012}, tunable or reconfigurable filtering \cite{Ebrahimi2016a,Azemi2013,Ebrahimi2015b}, and dual-band filtering response \cite{Ebrahimi2014,Salehi2008}. An essential factor in some spatial filtering applications such as radomes for narrowband antennas is a narrow band filtering response with sharp out-of-band rejection. However, there are a few research works investigating the design of FSSs with narrowband filtering responses. A major challenge in the design of such narrowband filters is the limited quality factor of the resonant elements used in the conventional FSSs \cite{Pous1991}. In order to achieve narrowband filtering responses with high selectivity, the quality factor of the constituting elements of the FSS should be increased by decreasing the overall size of the unit cells while keeping the resonance frequency unchanged. This is very challenging using the available PCB fabrication technologies \cite{Al-Joumayly2009}.

In this paper, we propose a miniaturized FSS unit cell that can achieve high quality factor resonance by appropriately designing its geometrical parameters. An equivalent circuit model is developed for modelling the electromagnetic response of the designed FSS and having a better insight to its operational principle. The developed equivalent circuit model is utilized for designing higher-order narrowband bandpass FSSs based on the standard filter theory. The operation principle and the design process of the proposed narrowband FSS will be discussed in the next sections.      

\begin{figure}[!t]
\centering
\includegraphics[width=2.2in]{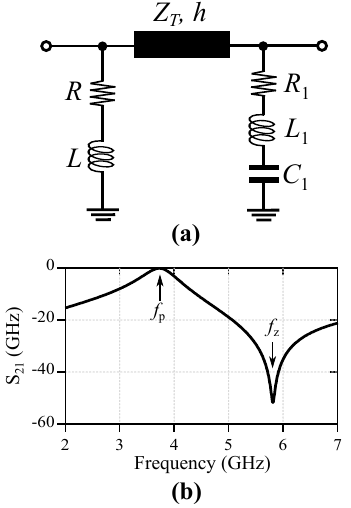}
\caption{(a) Basic circuit model of the proposed narrowband FSS. (b) A typical transmission response of the circuit.}
\label{Fig1}
\end{figure}

\section{FSS Structure and Operation Principle}
Fig.~\ref{Fig1}(a) presents the basic equivalent circuit model of the proposed FSS. The circuit is a hybrid resonator made of a series $L_1C_1$ resonator and a parallel inductance of $L$. The $R_1$ and $R$ model the losses in $L_1$ and $L$ inductors respectively. A typical transmission response of such a circuit is indicated in Fig.~\ref{Fig1}(b). As seen, the transmission response shows a passband in vicinity of a transmission zero. If the length of the transmission line stub ($h$) is much smaller than the wavelength in the operational frequency range, the frequency of the passband can be obtained as

\begin{equation}\label{Eq1}
f_\mathrm{p}=\dfrac{1}{2\pi\sqrt{(L+L_1)C_1}},
\end{equation}
and the transmission zero frequency is calculated as

\begin{equation}\label{Eq2}
f_\mathrm{z}=\dfrac{1}{2\pi\sqrt{L_1C_1}}.
\end{equation}

In addition, the quality factor of the passband is obtained as 

\begin{equation}\label{Eq3}
Q=\dfrac{1}{(R+R_1)}\sqrt{\dfrac{L+L_1}{C_1}}.
\end{equation}

\begin{figure}[!t]
\centering
\includegraphics[width=2.5in]{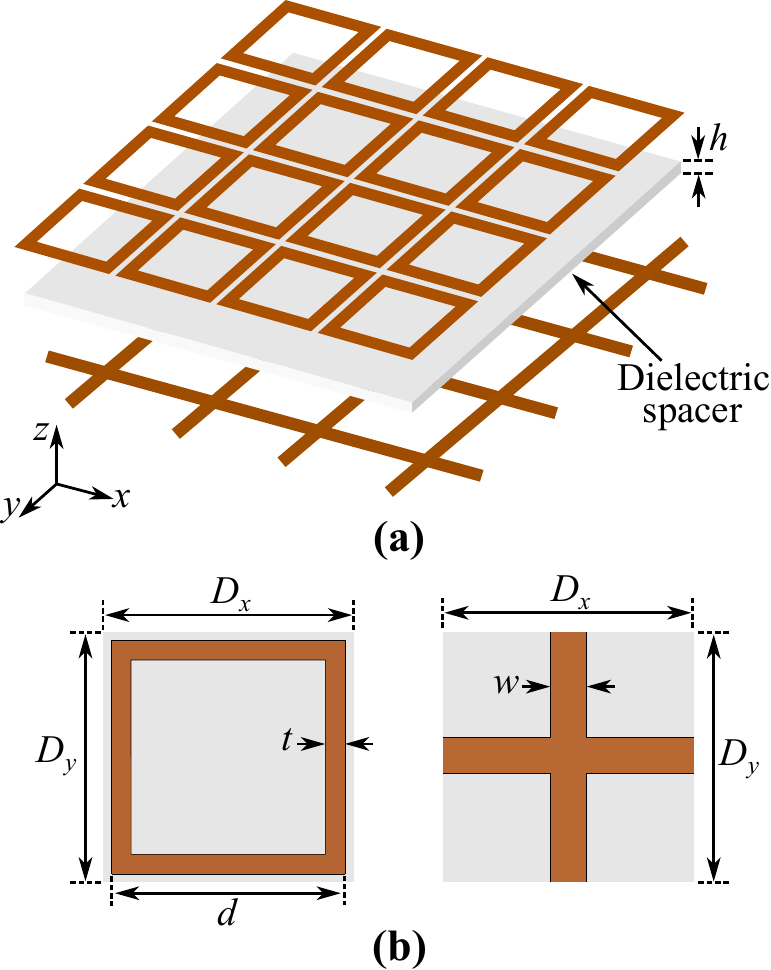}
\caption{(a) Three dimensional view of the proposed FSS. (b) Front and back layer unit cells of the FSS. The unit cell dimensions are: $D_x=~D_y=~10.2$~mm, $d=9.8$~mm, $t=0.4$~mm, and $h=0.254$~mm.}
\label{Fig2}
\end{figure}

The equivalent circuit model in Fig.~\ref{Fig1}(a) can be spatially implemented using the structure in Fig.~\ref{Fig2} that is a representation of the proposed FSS. In the proposed FSS, the series $L_1C_1$ resonator is implemented with the periodic arrangement of the square rings in the front layer, where $L_1$ is the equivalent inductance of the ring resonator sides, and $C_1$ is the equivalent capacitance between the adjacent square rings in the array. The parallel $L$ inductance is realized with an inductive wire grid in the back layer. The $R$ and $R_1$ resistances are the ohmic losses in the inductive wire grid and the ring resonators respectively. Furthermore, the short transmission line stub is implemented through a thin dielectric spacer separating the front and back layers metallic patterns, where $Z_T=Z_0/\sqrt{\epsilon_r}$. The $Z_0=377$~$\mathrm{\Omega}$ is the characteristic impedance of the free space and $\epsilon_r$ is the relative permittivity of the dielectric spacer. 

In order to have a narrow passband, the passband quality factor ($Q$) should be increased. Based on (\ref{Eq3}), the quality factor can be increased by decreasing ($R+R_1$), and increasing the equivalent inductance ($L+L_1$). For decreasing the equivalent resistance ($R+R_1$), the width of the wire grid in the back layer ($w$), or the width of the ring resonator arms ($t$) can be increased. However, this reduces the total inductance of ($L+L_1$) as well since in general, the thinner conductors result in larger inductances. But, it will be shown here that increasing the width of the wire grid in the back layer can improve the quality factor significantly. Now, consider the back layer wire grid. It is known that \cite{Ebrahimi2015} 

\begin{equation}\label{Eq4}
L \propto \ln \left( \dfrac{1}{\sin\dfrac{\pi w}{2D}} \right),
\end{equation}
where $D=D_x=D_y$ is the size of the FSS unit cell. On the other hand, we have 

\begin{equation}\label{Eq5}
R \propto \dfrac{1}{w}.
\end{equation}

Based on the above relations and (\ref{Eq3}), the increase in $w$ should increase $Q$ since the rate of changes in $R$ is larger than the $L$ variation due to the changes in $w$.

\begin{figure}[!t]
\centering
\includegraphics[width=3in]{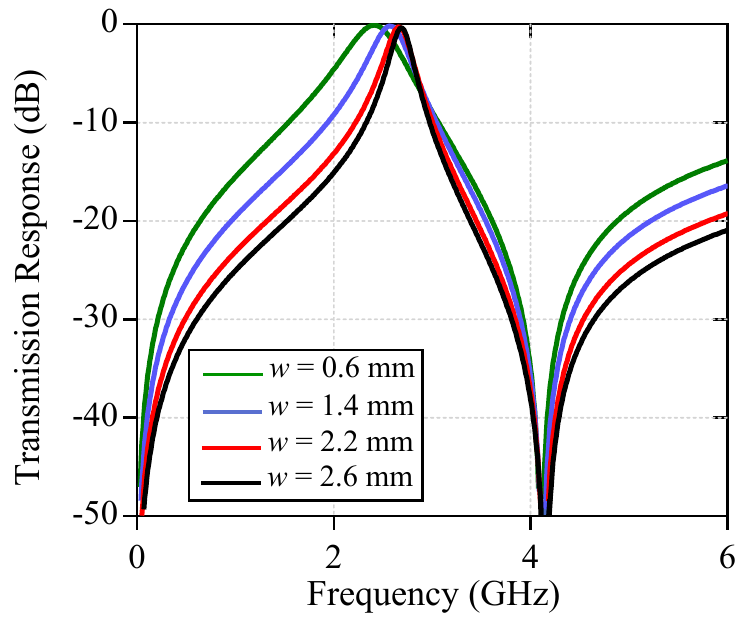}
\caption{Transmission responses of the FSS in Fig.~\ref{Fig2} for different values of $w$. }
\label{Fig3}
\end{figure}

\begin{figure}[!t]
\centering
\includegraphics[width=3in]{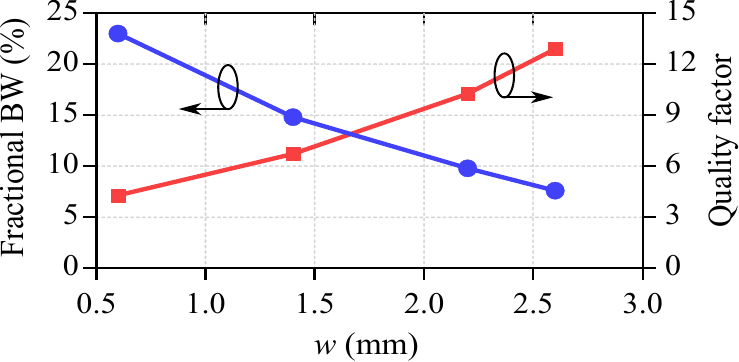}
\caption{Fractional bandwidth and the quality factor of the FSS passband as function of $w$. }
\label{Fig4}
\end{figure}

We have simulated a sample FSS with the geometrical parameters given in the caption of Fig.~\ref{Fig2} and different values of $w$ as a proof of concept. The used dielectric spacer in simulations is \textit{Rogers RO5880}, and the simulations are performed in \textit{CST Microwave Studio} software. The simulated transmission responses of the FSS for different values of $w$ are plotted in Fig.~\ref{Fig3}. Furthermore, the fractional bandwidth and the quality factor of the FSS passband are plotted versus $w$ in Fig.~\ref{Fig4}. As seen, we can achieve fractional bandwidths smaller than $10$\%  by increasing $w$ up to $2.6$~mm. It should be mentioned that the passband center frequency will be shifted up slightly by increasing $w$. Thus, in the design and optimization process the starting values of the dimensions should be chosen in way that the passband frequency is slightly smaller than the desired one. Then, it can be set to the desired frequency by an appropriate choice of $w$ based on the required bandwidth. 

\begin{figure}[!t]
\centering
\includegraphics[width=3in]{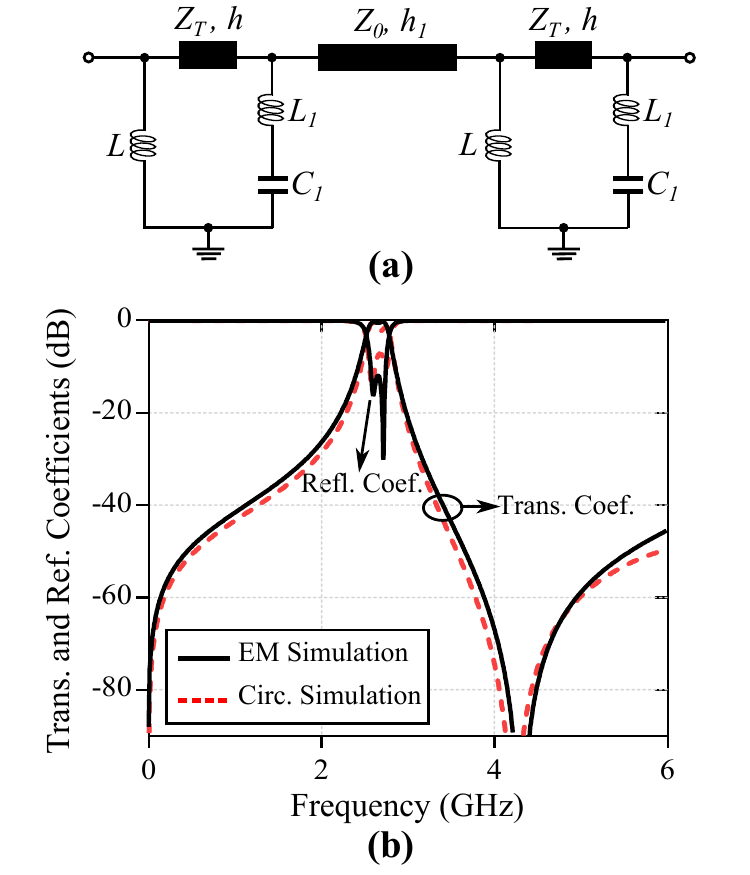}
\caption{(a) Equivalent circuit model of the second-order FSS made of two cascaded FSS layers in Fig.~\ref{Fig3}. (b) Electromagnetic and circuit model simulation results of the second-order FSS for normal incidence. The circuit parameters are: $L=2.85$~nH, $L_1=1.61$~nH, $C_1=0.6$~pF, $h=0.254$~mm, and $h_1=10$~mm. The geometrical dimensions of the FSS unit cell are given in Fig.~\ref{Fig3} and $w=2.6$~mm. }
\label{Fig5}
\end{figure}

\vspace{-2mm}
\section{Higher-order Filters}
\label{sec:Simulations}
The design of higher-order filtering responses is also possible based on the proposed FSS structure. To this end, the FSS layers can be cascaded by using quarter wavelength ($\lambda/4$) spacers acting as impedance converters between them \cite{Abadi2015}. For demonstration, a second-order FSS has been implemented here by cascading two layers of the FSS implemented in the previous section with a sub-wavelength air gap between them. The equivalent circuit model of the second-order FSS is presented in Fig.~\ref{Fig5}(a), where the transmission line with a length $h_1$ and a characteristic impedance of $Z_0=377~\Omega$ models the sub-wavelength air gap between the two FSS layers. A comparison between the full-wave electromagnetic and circuit model simulations of the designed second-order FSS is presented in Fig.~\ref{Fig5}(b). There is a good agreement between the circuit and the EM simulations verifying the presented circuit model. As seen, a narrow bandwidth of $8.5$\% is achieved around the center frequency of $2.7$~GHz.

\begin{figure}[!t]
\centering
\includegraphics[width=2.5in]{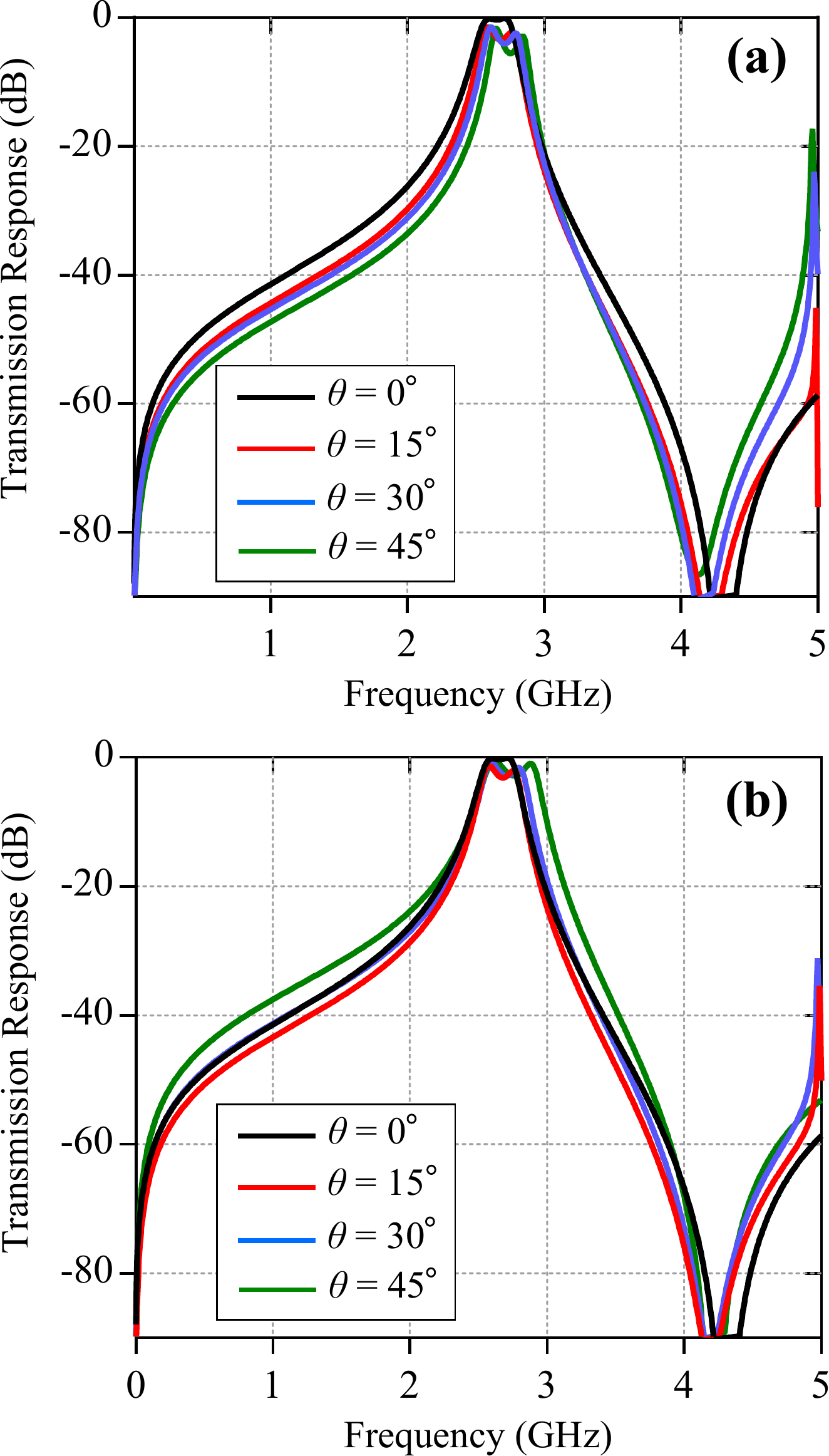}
\caption{Transmission responses of the designed second-order narrowband FSS under oblique incidence angles for (a) TE polarization, and (b) TM polarization of the electromagnetic field. }
\label{Fig6}
\end{figure}

The performance of the designed second-order FSS has been investigated for the oblique angles of the incidence as well. Fig.~\ref{Fig6} indicates the simulated transmission responses for the oblique angles for both of the TE and TM polarization. As seen, the FSS maintains its center frequency and the second-order characteristic over the incidences angles up to $45^\circ$. This is attributed to the miniaturized size of the FSS unit cell that is smaller than $\lambda/4$ at the center frequency of the filter.    
  
\vspace{-1mm}
\section{Conclusion}
A miniaturized element FSS has been proposed for designing narrowband bandpass FSSs in this paper. It is demonstrated that the first order FSS can achieve very narrowband responses with the fractional bandwidth smaller than $10$\% by a proper design of the unit cell geometry. The compatibility of the proposed first order FSS for designing higher order frequency responses has been demonstrated by designing an FSS with a second order frequency response. The full-wave electromagnetic simulations show a good stability of the second-order FSS frequency response over the incidence angles up to $45^\circ$.



\end{document}

\else
\fi

